\documentclass[aps,prb,
twocolumn, 
               a4paper,amsmath,amssymb,superscriptaddress]{revtex4-1}

\usepackage[utf8]{inputenc}
\usepackage[english]{babel}
\usepackage{graphicx}
\usepackage{feynmp}
\usepackage{bm}
\usepackage{color}

\newcommand{\e}{ {\rm e}}
\newcommand{\bk}{ \bm{k}}
\newcommand{\bq}{ \bm{q}}

\newcommand{\ep}{ \epsilon }
\newcommand{\tf}{ \tilde{f}}
\newcommand{\tn}{ \tilde{n} }
\newcommand{\g}{ \gamma }
\newcommand{\tthpm}{ \theta_0^{\pm}}
\newcommand{\tthmp}{ \theta_0^{\mp}}
\newcommand{\eno}{ \Gamma_0}
\newcommand{\tvs}{ \tilde{v}_s}
\newcommand{\ttt}{ \tilde{T}}
\newcommand{\tmu}{ \tilde{\mu}}
\newcommand{\norlam}{ \lambda/\lambda_0}

\begin{document}
\title{
Role of acoustic phonons in exotic conductivity of  two-dimensuional Dirac eletrons
}


\author{Yoshikazu \surname{Suzumura}}
\email{suzumura@s.phys.nagoya-u.ac.jp} 
 \affiliation{Department of Physics, Nagoya University,
             Nagoya 464-8602, Japan}            
\author{Masao \surname{Ogata}}
 \affiliation{Department of Physics, University of Tokyo, Bunkyo, Tokyo 113-0033, Japan}

\begin{abstract}
We examine the effect of acoustic phonon scattering on the conductivity of  two-dimensional Dirac electrons. 
 The temperature ($T$) dependence of the conductivity ($\sigma$) is calculated using the electron Green's function with damping by both the impurity ($\eno$) and phonon ($\Gamma_{\rm ph}$).
For zero or small doping, on which the present Rapid Communication focuses, $\sigma (T) $ increases and becomes almost constant due to the competition between the Dirac electrons and the phonon scattering. 
Such strange behavior of $\sigma (T)$ is ascribed to an exotic  mechanism of phonon scattering, whose momentum space  is strongly reduced 
 in the presence of a Dirac cone. 
For large doping, $\sigma$ decreases due to the interplay of 
 the  Fermi surface and the phonon.
 The unconventional $T$ dependence of the resistivity $\rho (=1/\sigma)$ for small doping is compared with   that of the experiment of  Dirac electrons
   in an organic conductor.
\end{abstract}

\maketitle

 Two-dimensional massless Dirac fermions are well known as a system exhibiting topological properties, e.g., the  quantum Hall effect in graphene monolayers \cite{Novoselov2005_Nature438}.
The linear dispersion of the energy spectrum with a Dirac cone gives 
 a topological anomaly associated  
with the Berry phase  \cite{Berry1984}, which can be  shown  
 using a spin rotation operator of the wave function \cite{Ando2005_JPSJ}. 
When the chemical potential is located at the energy of the Dirac point, i.e., zero doping, unconventional  properties appear in the transport phenomena.   
While the density of states vanishes linearly at the energy of the Dirac point,
 the conductivity even in the limit of weak impurities   
 displays not a zero-gap semiconductor 
 but a metallic behavior. 
In fact, the conductivity at zero doping is known as
 a minimum conductivity with 
a universal constant  \cite{Novoselov2005_Nature438,Fradkin1986,Ando2005_JPSJ,
Ziegler2007,Castro2009_RMP}. 
 This quantum conductivity comes from the singularity of the Dirac cone, 
while  the conductivity treated classically by the Boltzmann equation is reduced to zero at  zero doping \cite{Castro2009_RMP}.

Thus, the conductivity is one of the most fundamental phenomena and  has been shown to have very peculiar behaviors in bulk systems of two molecular conductors \cite{Kajita_JPSJ2014,Kato_JACS}.
One is the two-dimensional organic conductor [BEDT-TTF=bis(ethylenedithio)tetrathiafulvalene],  in which a zero-gap state with a tilted Dirac cone was found using a tight-binding model 
 \cite{Katayama2006_JPSJ75,Kondo2005}.
The other is a single-component molecular conductor under  high pressures, which shows a three-dimensional Dirac electron with a nodal line semimetal  
\cite{Kato2017_JPSJ}. 
 These conductors share a common feature for the temperature dependence of the conductivity. 
 It is known theoretically that the conductivity of two-dimensional Dirac electrons  at absolute zero temperature becomes a  constant even 
 within the self-consistent  Born approximation \cite{Ando1998}
 and increases with increasing temperature. \cite{Katayama2006_cond,Kobayashi2008}
 However, the calculation of the temperature dependence \cite{Katayama2006_cond,Kobayashi2008}  was performed using the damping  only by impurities 
 and  
 without taking account of a mechanism of suppressing the conductivity 
 by phonons at finite temperature.
 Indeed, the resistivity observed in these conductors is almost constant with increasing temperature. 
In spite of the strong temperature dependence of the Hall coefficient \cite{Kajita_JPSJ2014}, the constant resistivity has been regarded  as  evidence of 
 the presence of  a Dirac electron 
\cite{Tajima2007_EPL,Kato_JACS}.

In this Rapid Communication,  the constant resistivity of the two-dimensional  Dirac electrons of the organic conductor  
 is studied by examining  the scattering  by   acoustic phonons in addition to impurity scattering. 
The temperature dependence of resistivity of the Dirac electrons with phonon scattering  has been studied for the doped graphene using the Boltzmann equation \cite{Sarma2008,Sarma2011}.  
 They show that the scattering rate depends linearly on the temperature ($T$) at high temperatures and is proportional to $T^4$ at low temperatures. 
 This $T$ dependence for large doping is reasonable due to the existence of the  Fermi surface. 
 However, such mechanism of scattering is invalid for the zero or small doping due to the absence of a robust Fermi surface. Here, we demonstrate that the unconventional $T$ dependence of resistivity originates from the fact that the momentum space of the phonon scattering is strongly reduced at zero or small doping. 
For such cases, it is important to carry out the linear response calculation based on the Kubo formula, not in the Boltzmann equation.  We will show that the interplay of such phonon scattering and the Dirac cone is  crucial  to comprehend the unconventional  temperature dependence of the conductivity 
 and resistivity of  Dirac electrons.

We consider a two-dimensional electron-phonon ($e$-$p$) system given by 
\begin{equation}
H= H_0 + H_{int} + H_{imp} \; , 
\label{eq:H}
\end{equation}
which is the   Fr\"ohlich Hamiltonan 
\cite{Frohlich}  
   applied to the  Dirac electron system.
$H_0$ is expressed as 
\begin{equation}
  H_0 = \sum_{\bk} \sum_{\g = \pm} \ep_{\bk, \g} a_{\bk, \g}^\dagger a_{\bk,\g}
     + \sum_{\bq} \omega_{\bq} b_{\bq}^{\dagger} b_{\bq} \; , 
\label{eq:H_0}
\end{equation}
 where 
 $\ep_{\bk, \g} = \g v k$, 
 $\omega_{\bq} = v_s q$,
with $k=|\bk|$ and  $q=|\bq|$, and $v \gg v_s$. 
The lattice constant is taken as unity. 
$\g = +$ $(= -)$ denotes a conduction (valence) band with a  Dirac cone. 
$a_{\bk,\g}$ and $b_{\bq}$ are annihilation operators of the electron with a wave vector $\bk$ of the  $\g$ band and acoustic phonon with a vector $\bq$. 
The first term of Eq.~(\ref{eq:H_0})  is  obtained by diagonalizing 
 a $2$ $\times$ $2$ isotropic model which is simplified compared with 
the effective model for the  Dirac electron in the organic conductor 
\cite{Kobayashi2007}, 
  but contains a  minimum ingredient  to comprehend the effect of phonon 
 on the Dirac electron.
 The electron-phonon ($e$-$p$) interaction  given by $H_{\rm int}$  is expressed as 
\begin{equation}
 H_{int} = \sum_{\bk, \g} \sum_{\bq}
   \alpha_{\bq} a_{\bk + \bq, \g}^\dagger a_{\bk, \g} \phi_{\bq} \; ,
\label{eq:H_int}
\end{equation}
where 
 $\phi_{\bq} = b_{\bq} + b_{-\bq}^{\dagger}$.
  We introduce  a coupling constant $\lambda = |\alpha_{\bq}|^2/\omega_{\bq}$ 
  which becomes  independent of $|\bq|$  for small $|\bq|$. 
The $e$-$p$ scattering is considered 
 within  the same band (i.e., intraband) 
  due to the energy conservation with $v \gg v_s$. 
The third term of Eq.~(\ref{eq:H}), $H_{\rm imp}$, denotes a normal  impurity 
 scattering, which is introduced to avoid infinite conductivity 
in the presence of only the $e$-$p$ interaction 
\cite{Holstein1964}. 
We take $k_{\rm B} = \hbar$ = 1.

First,  we note a relation between the chemical potential $\mu$ and 
 $\mu_0$  where  $n = \mu_0^2/2\pi v^2$  and  the doping rate $n$  
    is calculated   
  from the density of states (DOS),
 $D(\omega) = \sum_{\bk} \delta(\omega - \ep_{\bk}) = |\omega| /( 2\pi v^2)$. 
  Using $y = (\omega-\mu)/T$ 
 and  $\Theta(x)$ (= 1  for $x>0$  and 0   otherwise),
 the self-consistency  equation for $\mu$ is given by 
$\int_{-\infty}^{\infty } d \; \omega \;$ 
$D(\omega) [ 1/({\rm e}^y + 1) - \Theta (\mu_0 - \omega) ]
 = 0$, 
which is rewritten as $(\mu/\mu_0)^2 =  1 - B(\mu,\mu_0,T)$ with 
\begin{eqnarray}
  B = 2 (T/\mu_0)^2  \int_{0}^{\infty} d \; y\;
  \frac{y+ \mu/T - |y - \mu/T|}{\e^y + 1} \; .
\label{Eq:mu_SCE}
 \end{eqnarray}  
When  the doping is zero ($\mu_0=0$),
$\mu=0$ for arbitrary $T$.
Equation (\ref{Eq:mu_SCE}) shows  that 
 $\mu/\mu_0$ is determined only  by   $T/\mu_0$.  
The asymptotic form is given by   $\mu/\mu_0 \simeq 1 - (\pi^2/6)(T/ \mu_0)^2$ 
 for $T/\mu_0 < 0.3$, and 
$\mu/\mu_0 \simeq (\mu_0/T) /(4 \; {\rm ln} 2)$ for $1 < T/\mu_0$. 
This form reproduces well the exact  result within the visible scale 
  except for  the intermediate region, e.g.,    
  $\mu/\mu_0 = $ 0.760, 0.575, and 0.444 for 
   $T/\mu_0 = $ 0.4, 0.6,  and 0.8, respectively.

 Using a self-energy of the electron Green's function,
 the damping of an electron by a phonon   is obtained as 
\cite{Abrikosov} 
\begin{eqnarray}
 & & \Sigma_\g (\bk, n)  =  T \sum_m \sum_{\bq}\; |\alpha_q|^2 
            \nonumber \\
 & &\times   \frac{1}{i \omega_{n+m} - \xi_{\bk+\bq, \g}} 
     \;  \frac{2 \omega_{\bq}}{\omega_{m}^2 + \omega_{\bq}^2} \; , 
 \label{eq:self_energy}
  \end{eqnarray} 
which is a product of the electron and phonon Green's functions. 
$\omega_n=  (2n+1)\pi T$, $\omega_{m}=2\pi m T$, with $n$ and $m$ being integers. $\xi_{\bk, \g} = \ep_{\bk, \g} - \mu$, where  $\mu$ is a chemical potential measured  
 from the energy of the Dirac point.  
 In Eq.~(\ref{eq:self_energy}),
   the Green's function without interactions  is used 
   in the sense of the perturbational method. 
After the analytical continuation $i \omega_n \rightarrow \omega + i \delta$       with $\delta = + 0$, 
 the imaginary part is calculated as 
\begin{eqnarray}
& &  {\rm Im} \Sigma_\g (\bk, \omega + i \delta) = 
  -  \Sigma'_\g (\bk, \omega) = 
   - \pi \sum_{\bq} |\alpha_q|^2 
      \nonumber \\
 & &  \times   \left\{  
     \delta ( \omega + \omega_{\bq} - \xi_{\bk+\bq, \g}) \times
      \left[ n_{\bq}+ f(\xi_{\bk+\bq, \g}) \right] 
\right.
  \nonumber \\ 
 & &  + 
 \left. \delta ( \omega - \omega_{\bq}  - \xi_{\bk+\bq, \g}) \times
 \left[ n_{\bq}+ 1 - f(\xi_{\bk+\bq, \g}) \right]
 \right\}  \; , 
 \nonumber \\ 
 \label{eq:imag_self_energy}
 \end{eqnarray} 
where $n_{\bq} = 1/(\exp[\omega_{\bq}/T]-1)$ and  $f(x) = 1/(\e^{x/T} + 1)$. 
Note that  $\Sigma'_\g (\bk, \omega)$ reduces to zero at $T=0$, since 
 the phonon is absent and the excitation between the electron and hole becomes zero  due to $n_{\bq}$ and $f(\xi_{\bk+\bq, \g})$, respectively.  

The Green's function with  the damping of 
 both the impurity and phonon scattering is  given as 
\begin{eqnarray}
 G_\g(\bk, i \omega_n)^{-1}  =  
 i \omega_n + \ep_{\bk,\g}- \mu 
  + i \Gamma_{\g}  
  \; ,
 \label{eq:green_function} 
  \end{eqnarray} 
 where 
$\Gamma_{\g}  =  \Gamma_0 + \Gamma_{\rm ph}^{\g}$ and 
$\Gamma_{\rm ph}^{\g} = \Sigma'_\g (\bk, \g v k - \mu)$. 
 $\Gamma_0$,  which is the damping by impurity scattering,  
  is taken as  a parameter to scale the energy. 
 Note that Eq.~(\ref{eq:imag_self_energy}) can be improved  
 by introducing 
 $\eno$  in Eq.~(\ref{eq:self_energy}) for the electron Green's function. 
This  gives the  Lorentzian function instead of the 
 $\delta$ function  in Eq.~(\ref{eq:imag_self_energy}) 
  and the resultant reduction of   $\Sigma'_\g (\bk, \omega)$ 
 is large (small) for  $T < \eno$  ($T > \eno$).

Using a linear response theory and Eq.~(\ref{eq:green_function}),
  the conductivity  $\sigma_{xx}$
 along the $x$ direction 
is calculated from 
\begin{eqnarray}
 \sigma_{xx} =  \sigma (T) =  i \lim_{\omega \rightarrow 0} 
  \frac{Q(\omega) - Q(0)}{\omega} 
  \; ,
 \label{eq:green_function_response_1}
\end{eqnarray}
\begin{eqnarray}
 Q(i \omega_m)  =  2 e^2 T \sum_{n} \sum_{\bk} \sum_{\g, \g'}
|v_{\g,\g'}(\bk)|^2  \Pi_{\g, \g'}(n,m,\bk) \; ,
  \label{eq:green_function_response_2}
 \end{eqnarray}
 where $\Pi_{\g, \g'}(n,m,\bk) = 
   G_{\g'}(\bk, i \omega_{n+m}) G_\g(\bk, i \omega_n)$,  
 and $Q(\omega) = Q( i \omega_m)$ at 
  $ i \omega_m \rightarrow  \omega + i \delta$. 
 The factor 2 comes from the spin.
The velocity matrices are given by 
  $|v_{\g,\g'}(\bk)|^2 = v^2(k_x/k)^2$  for $\g = \g'$ (intra-band)
 and $v^2(k_y/k)^2$ for  $\g = -\g'$ (inter-band) where $\bk =(k_x, k_y)$.

From Eq.~(\ref{eq:green_function_response_2}) 
  with a vertex correction  included in $\Gamma_{\pm}$,  
 the conductivity normalized by $\sigma_{xx}^0  = e^2/2\pi^2$ per spin 
is given as follows 
 \cite{Katayama2006_cond}, 
\begin{eqnarray}
 \frac{\sigma(T)}{\sigma_{xx}^0} & = &
   \int_{-\infty}^{\infty} 
    d \;z  \; \left(- \frac{\partial f(z)}{\partial z}\right) F(z) \; ,
\label{eq:sigma_total} 
\end{eqnarray}
\begin{eqnarray}
 & &F(z) =  \int_{0}^{\infty}  d \; \eta \; \eta \; 
       \left( \frac{\Gamma_+/\eno}{(z/\eno-\eta+\mu/\eno)^2  
    + (\Gamma_+/\eno)^2} \right.
   \nonumber \\
    & &   + \left.  \frac{\Gamma_-/\eno}{(z/\eno + \eta + \mu/\eno
     )^2+(\Gamma_-/\eno)^2} \right)^2 \; ,
\label{eq:Fz_total_scale} 
 \\
 & & \frac{ \Gamma_\pm}{\eno} -1 
    = g_{\pm}(\mu/\Gamma_0,\tvs,T/\Gamma_0, \eta) =  \lambda K \eta^2  h_\pm 
 \; ,
\label{eq_gamma_pm_scale} 
\end{eqnarray}
where $g_{\pm} = \Gamma_{\rm ph}^{\pm}/\eno$,
  $\eta = \xi/\eno$, $\xi = vk$  and  $K = (v_s/v) \eno /(2 \pi v^2)$.
As typical values,  we will take 
 $\lambda = \lambda_0  = $ 0.031 eV, $v_s/v = $ 0.05, and 
    $\ v \sim$ 0.05 eV for organic conductors \cite{Katayama_EPJ},  
 which lead  to $\lambda_0 K =  10^{-4}$  for $\eno = $ 0.001 eV.
The dimensionless $e$-$p$ coupling constant is given by
   $\lambda_0 / (2 \pi v) \simeq  0.1$. 
Note that $g_{\pm}$ at high temperature  of $T \gg v_s q$ 
 is almost independent of $v_s$ for fixed $\lambda$ .  
The quantity $h_\pm$  is given by 
\begin{eqnarray}
& &  h_\pm \equiv 
 \int_0^{2/(1 \mp \tvs)}  d \;y  \; y \; 
    \frac{\sqrt{1 + y^2 +2 y \cos \tthpm}}{ \sin \tthpm} 
   \nonumber \\
 & &  \times (1 -\cos \theta^{\pm}) 
    [(1-\tf_1^\pm)\tn + \tf_1^\pm(1+\tn)] \nonumber \\
  && +
  \int_0^{2/(1 \pm \tvs)}  d \;y  \; y \; 
     \frac{\sqrt{1 + y^2 +2 y \cos \tthmp}}{ \sin \tthmp}
  \nonumber \\
& & \times    (1 -\cos \theta^{\mp}) 
     [(1-\tf_2^\pm)(\tn+1) + \tf_2^\pm \tn] 
\; ,
\nonumber \\
\label{eq_g_pm} 
\end{eqnarray}
  with  $\tn= (\e^{\tvs y/\ttt}-1)^{-1}$, 
 $y=q/k$,  $\ttt=T/\xi$, $\tvs = v_s/v$, $\tmu = \mu/\xi$,
 $\tf_1^\pm = 1/(\e^{(\pm 1 + \tvs y - \tmu)/\ttt}+1)$, 
 $\tf_2^\pm = 1/(\e^{(\pm 1 - \tvs y - \tmu)/\ttt}+1)$, 
$\cos \tthpm = (\tvs^2 -1 )y/2 \pm \tvs$, and $\tthpm> 0$.
The condition  of  
 $ q/k < 2/(1 \pm \tvs)$  in  Eq.~(\ref{eq_g_pm}),
  which originates from the energy conservation with 
  the Dirac cone  in Eq.~(\ref{eq:imag_self_energy}) 
 gives  $vq$ (not $v_s q$ ) bounded by $T$ 
  in the calculation of $\sigma (T)$. 
Thus  $g_{\pm}$ for  small doping  is reduced 
 by a factor $\tvs (\ll 1)$  
 compared with the conventional case with a robust Fermi surface. 
 In Eq.~(\ref{eq_g_pm}), a factor $(1 - \cos \theta^{\pm})$ 
 has been introduced 
 as a vertex correction \cite{note},  
 where 
$\cos \theta^{\pm} =(1 + y \cos \tthpm)/\sqrt{ 1 + y^2 + 2 y \cos \tthpm}$. 
Note that $\sigma$ is an even function with respect to $\mu$. 

\begin{figure}
  \centering
\includegraphics[width=6cm]{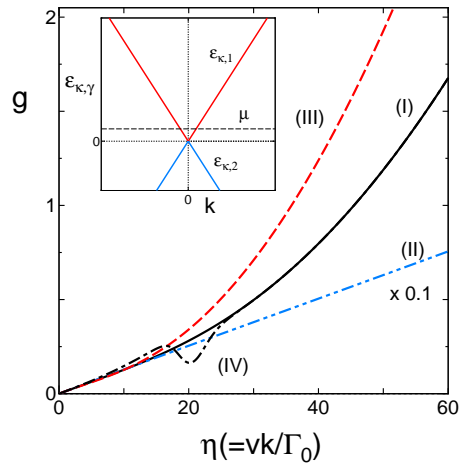}
     \caption{(Color online)
The damping given by the e-p scattering,  $g (= g_+)$ 
 derived in Eq.~(\ref{eq_gamma_pm_scale})
 as a function of $\eta (= vk/\eno)$ for 
 $(\mu/\eno, v_s/v, T/\eno) 
 = $  $(0,0.05,1)$ (I), $(0,0.05,10)$ (II), $(0,0.1,1)$ (III), 
 and $(20,0.05,1)$ (IV), respectively. 
It is found that $g \propto \eta T$  
 for  $\eta < 10$ as in Eq.~(\ref{eq:eq14}). 
 A dip exists around $\eta=\mu/\eno$ 
 as seen in case IV for   $\mu/\eno = $ 20.  
The inset denotes the schematic energy dispersion of Dirac cones 
 for the conduction $(\ep_{\bk,1})$ and valence $(\ep_{\bk,2})$ bands where
 $\mu$ denotes the chemical potential close to zero doping.
}
\label{fig:fig1}
\end{figure}

 Using scaled quantities   $T/\eno$ and  $\mu/\eno$,  
 we examine numerically 
  the normalized conductivity $\sigma(T)$ of Eq.~(\ref{eq:sigma_total}).
 Figure \ref{fig:fig1} shows the normalized damping $g = g_+$ 
 as a function of $\eta (= vk/\eno)$ for 
  $(\mu/\eno, v_s/v, T/\eno) 
 = $ $(0,0.05,1)$ (I), $(0,0.05,10)$ (II), $(0,0.1,1)$ (III), 
 and $(20,0.05,1)$ (IV), respectively. 
  The inset denotes the energy dispersion of the Dirac cone and 
 chemical potential $\mu$ close to zero doping. 
 When $\mu = 0$, $g_+ = g_- = g$, and 
 the lines (I), (II) and (III) for  $\eta < 10$ 
  suggest   a formula   
\begin{eqnarray}
 &&  g \simeq 1.25 \times 10^{-2} (\lambda/\lambda_0) (T/\eno) \eta  
  \; ,
 \label{eq:eq14}
 \end{eqnarray}
where  $h_{\pm} \simeq 6.25 \times T/(vk \tvs)$ 
  for $\tvs=0.05$. 
Equation (\ref{eq:eq14}) comes 
    from  the classical  phonon (i.e., $v_s q < T$)
since $\eta (=vk/\eno) < 10$ in Fig.~\ref{fig:fig1} corresponds to 
   $v_sq/\eno < v_sk/\eno < 1  =T/\eno$.  
  The validity of Eq.~(\ref{eq:eq14}) is shown   
   in the following calculation of the conductivity.
For $T < vk$, the deviation of $g$ from the linear dependence of $\eta$     is seen but does not contribute to $\sigma$, 
which is  obtained  for $vk \simeq T$. 
 For  $\mu/\eno \not= 0$,   
 $g_+$ exhibits a dip around $\eta \simeq \mu/\eno$ as shown by the line (IV).  Such an effect 
    becomes noticeable   with a further increase of  $\mu/\eno$, where 
  the behavior of the quantum phonon is expected at low temperatures.

\begin{figure}
  \centering
\includegraphics[width=6cm]{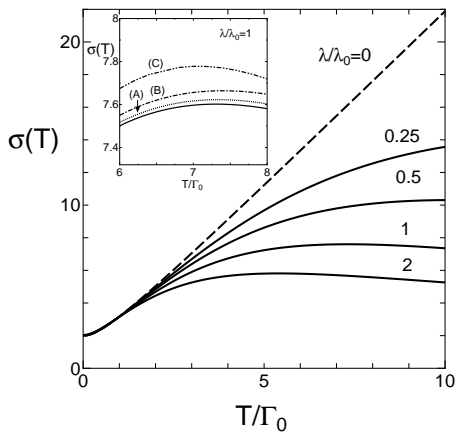}
     \caption{(Color online)
Normalized conductivity as a function of $T/\Gamma_0$ 
 for $\lambda/\lambda_0 = $ 2, 1, 0.5,and  0.25 
 with  $\lambda_0 = $ 0.0625 eV 
 where the corresponding dimensionless  coupling  constant is given by 
 $\lambda/(2\pi v) = $ 0.2, 0.1, 0.05, and  0.025, respectively.
The dashed line  corresponds to $\lambda = 0$
 (i.e., in the absence  of $e$-$p$ scattering).
The inset ($\lambda/\lambda_0 = 1$) denotes 
 a behavior close to the maximum  of $\sigma$, where  
 the solid line corresponds to  the main figure, 
 the dotted line (A) is obtained without a vertex correction, 
  the dotted-dashed line (B) is caculated 
  by substituting   Eq.~(\ref{eq:eq14}) 
 into  Eq.~(\ref{eq_gamma_pm_scale}), and 
  the double-dotted-solid line (C) is calculated  using Eq.~(\ref{Eq_app_sigma1}). 
}
\label{fig:fig2}
\end{figure}

 In Fig.~\ref{fig:fig2}, 
  $\sigma (T)$  is examined 
 for  $\mu = $ 0 (i.e. zero doping) and $v_s/v = 0.05$ 
 with some choices of $\lambda$.    
 The case of $\lambda = 0$ is also shown by the dashed line where 
    $\sigma (T) \sim 2 T/\eno $  for large $T/\eno$ 
   is understood from   $\sigma(T) \propto D(T)$ 
    with $D(T) = T /( 2\pi v^2)$.
 At low temperature, where 
  Eq.~(\ref{eq_gamma_pm_scale}) is smaller than 1, 
  the dominant contribution in 
 Eq.~(\ref{eq:Fz_total_scale}) comes from $\xi \sim \eno$ for small $T/\eno$  
 and   $\xi \simeq T$ for $T > \eno$ 
due to a factor 
 $- \partial f(z)/ \partial z$ in Eq.~(\ref{eq:sigma_total}), leading to   
 the increase of  $\sigma (T)$.
 However, 
 $g_{\pm}$ in Eq.~(\ref{eq_gamma_pm_scale}) becomes larger than 1 
 for $T$   above a crossover temperature 
 at which $\sigma (T)$ takes a maximum. 
 At higher temperatures, the  dominant contribution in Eq.~(\ref{eq:Fz_total_scale}) is obtained for  $\xi/\eno \simeq \lambda K \eta^2  g_\pm$.
 Furthermore, the  maximum of $\sigma(T)$  is large  
 for  small $\norlam$.
  In the inset of Fig.~\ref{fig:fig2}, several results 
   are compared with the main figure for $\norlam = $ 1 
(solid line) in a magnified scale. 
 The dotted-dashed line is obtained  
  by substituting   Eq.~(\ref{eq:eq14}) 
   into  Eq.~(\ref{eq_gamma_pm_scale}), and 
  the double-dotted-solid line is calculated  using Eq.~(\ref{Eq_app_sigma1}) as shown later.
The dotted line is obtained without a vertex correction.  
  These differences are almost invisible in the scale of the main figure. 
Thus the  effect of the vertex correction of the $e$-$p$ interaction 
is negligibly small, 
 which  is given by a factor $(1 - \cos \theta^{\pm})$ in Eq.~(\ref{eq_g_pm}).
Note that $\sigma (T)$ coincides quite well with 
 that obtained using  Eq.~(\ref{eq:eq14}). 
  Further,  the intraband contribution becomes  much larger than 
  that of the  interband  for $T/\eno  > 1$  although 
 both are the same at $T=0$ \cite{Suzumura2014}.   
 In fact,  
    the former exhibits a maximum while the latter stays almost constant with 
increasing $T$. 

\begin{figure}
  \centering
\includegraphics[width=6cm]{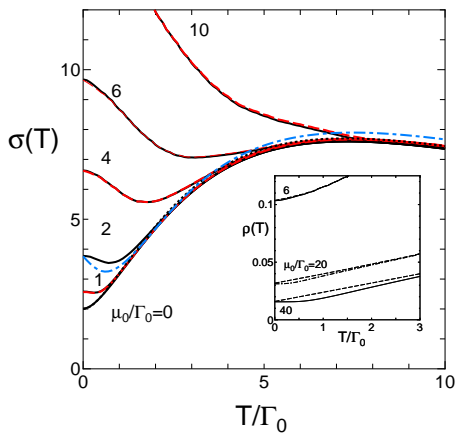}
     \caption{(Color online)
Normalized conductivity (solid line) 
  as a function of $T/\Gamma_0$ 
 with $v_s/v = 0.05$ 
 for $\mu_0/\Gamma_0 = $ 0, 1,  2, 4, 6, and 10. 
The  dashed line is obtained using    
 Eq.(\ref{eq:eq14}), where  the deviation from the 
 solid line is invisible.
The dotted-dashed line is shown as an example of 
  $\sigma (T)$ when we assume that $\mu/\eno = 2  -  T/\eno$ (see details 
 in the text).
In the inset, the resistivity $\rho (T)$ $[= 1/\sigma (T)]$ 
 for $\mu_0/\eno = $ 6, 20, and 40 
 is compared with that of the classical approximation 
 by Eq.~(\ref{eq:eq14}) (dashed line).
}
\label{fig:fig3}
\end{figure}

In Fig.~\ref{fig:fig3}, $\sigma (T)$ is shown 
 with several choices of $\mu_0/\eno$ where the corresponding  
  $\mu$  is calculated self-consistently using Eq.~(\ref{Eq:mu_SCE}). 
 At $T = $ 0,  
 $\sigma (0)$ increases monotonically with increasing $\mu$  
 due to   $D(\mu)$. 
 For small $\mu_0/\eno$, $\sigma (T)$ takes a minimum and a  maximum 
 while $\sigma (T)$ decreases monotonically for large $\mu_0/\eno$.
 A minimum of $\sigma$ at lower temperatures 
 for $\mu_0/\eno =$ 1, 2, 4, and 6 
  comes from the $T$ dependence of $\mu/\mu_0$ which decreases 
  rapidly for $\mu_0/\eno > 0.5$. 
Such a  minimum due to the chemical potential was also shown   
  for ballistic graphene \cite{Muller2009}.  
The maximum at higher temperatures is ascribed to  the following 
 competition. 
 The conductivity  at low temperatures is enhanced  
    by a thermaly excited carrier due to  the DOS of  the Dirac cone.
 At high temperatures, the effect of the damping $\Gamma_{\rm ph}$   
  by the phonon scattering 
  becomes compatible  with that  of the DOS ($\propto T$) 
     since the average of $\Gamma$ in  $\sigma (T)$ 
      gives $ \langle \Gamma \rangle \propto T^2$ from Eq.~(\ref{eq:eq14}).
   Thus the maximum of $\sigma(T)$  at  small doping originates from 
     the competition between   the  Dirac cone and  the phonon. 
     At high temperatures, 
    the difference in $\sigma (T)$ between $\mu_0/\eno = 0$ 
     and $\mu_0/\eno \not= 0$ becomes  small, 
      since  the effect of $T$   becomes  larger than that of $\mu$. 
  The asymptotic behavior of $\sigma(T)$ for large $T (\gg \eno)$ is given by 
    2, corresponding to a universal conductance.  
 In Fig.~\ref{fig:fig3}, we also show the dashed line  
     which is obtained using a formula of Eq.~(\ref{eq:eq14}).
 A good coincidence between two results suggests 
   that the $e$-$p$ scattering is determined by the classical phonon 
      not only for large $T/\eno$  but also for  small $T/\eno$. 

We mention  behaviors for large doping. With increasing $\mu_0/\eno$,     
 $\sigma(0)$ increases due to the increase of DOS, i.e., $D(\mu)$,  
 and the maximum in $\sigma (T)$ disappears.
In this case, the  scattering  at low temperatures is determined 
 by the quantum phonon since the condition $ \omega_{\bm{q}} > T$ 
  is allowed  for $\mu_0/\eno \gg 1$.
The case of such doping is shown 
 in the inset of Fig.~\ref{fig:fig3}, where 
 the resistivity $\rho ( =1 /\sigma)$
   at low temperature 
  is compared with that of the classical phonon  calculated  
 by Eq.~(\ref{eq:eq14}) (dashed line).
  The $T$-linear dependence of the classical phonon  
 is replaced by the $T^4$ dependence of  
   the quantum phonon ($\mu_0/\eno =$ 40) which comes from 
   the vertex correction 
  with a factor $(1-\cos \theta^{\pm})$ in Eq.~(\ref{eq_g_pm}). 
 Such behavior  for the large $\mu_0/\eno$,
  which  comes from  the reduction of $g_+$ in 
  Eq.~(\ref{eq:Fz_total_scale}) at $vk/\eno \sim \mu/\eno$  [see the line (IV) 
 in Fig.~\ref{fig:fig1}],  
  leads to conventional phonon damping in the presence of  
 a robust Fermi surface \cite{Sarma2008}.

We estimate the maximum of $\sigma (T)$ obtained in Fig.~\ref{fig:fig2} 
 using Eq.~(\ref{eq:eq14}) in which $\eta$ can be replaced by 
 $ \sim T/\eno$ 
from Eq.~(\ref{eq:Fz_total_scale}).
 Using $g = 0.02 (T/\eno)^2$  for $\lambda /\lambda_0 = 1$
 and $\sigma(T) = 1.1 (T/\eno) $ for the dashed line ($\lambda =$  0) 
 except for small  
 $T /\eno $ in Fig.~\ref{fig:fig2},  
 we obtain $\sigma(T)$ with  $\mu = 0$ and $\eno = 0.001$  as  
\begin{eqnarray}
 \sigma (T) \simeq \frac{1.1 \times (T/\eno)}
  {1 +0.02 \times (\lambda/\lambda_0)(T/\eno)^2} \; ,
\label{Eq_app_sigma1}
\end{eqnarray}
which,   within the visible scale,  reproduces the  line   
 for  $\lambda/\lambda_0 =$ 1  and  $T/\eno >4$ in Fig.~\ref{fig:fig2}. 
Note that $\Gamma_{\rm ph} = g \eno  [\propto (\lambda/\lambda_0) T^2]$ for the classical phonon is independent of $\eno$. 
Using Eq.~(\ref{Eq_app_sigma1}),   
the conductivity for arbitrary $\eno$ is given by 
$\sigma (T) \simeq 2.2 \times  T/ (\eno  + \Gamma_{\rm ph}),$
where $\Gamma_{\rm ph} = 20 (\lambda/\lambda_0)T^2$
 $(> \eno)$.
Thus, the maximum of $\sigma(T)$  
is estimated as 
$\simeq 7.7 /\sqrt{(\lambda/\lambda_0) \eno \times 10^{3}} $ 
 at $T \simeq 7 \sqrt{10^3 \eno/(\lambda/\lambda_0)}$. 
 
Here, we discuss the $T$ dependence of $\mu$, which was used to 
 analyze the experiment \cite{Kobayashi2008}. 
For simplicity, we calculate $\sigma (T)$  
   with a choice of  $\mu/\eno = 2 - T/\eno$, 
 which is qualitatively similar to 
 that taken  to explain 
  the  $T$ dependence of 
the  carrier density \cite{Tajima2007_EPL} and the Hall coefficient \cite{Kobayashi2008}. 
The result  is shown  
 by the dotted-dashed line in Fig.~\ref{fig:fig3}.  With increasing $T$, the crossover from electron doping to  hole doping occurs and the minimum of 
$\sigma (T)$  exists at lower temperature than $\mu(T)=0$.    
Such  $\mu (T)$  also exhibits a behavior that 
  $\sigma (T)$ is almost $T$ independent at high temperatures.

Further, using  Eq.~(\ref{eq:imag_self_energy}),  we estimate  
  the real part of the self-energy, 
 where  $\Sigma_{\gamma}(\bk, \omega) = \Sigma_1 + i \Sigma_2$, 
   $\Sigma_1 = {\rm Re}\Sigma_{\gamma}$, and 
     $\Sigma_2 = {\rm Im} \Sigma_{\gamma}$. 
Using the Kramers-Kronig relation, 
$\Sigma_1$ is calculated from $\Sigma_2$ as 
\begin{equation}
 \Sigma_1(\bk,\omega) 
   = \frac{1}{\pi} \int_{-\omega_0}^{\omega_0} {\rm d} \; z \; 
 \frac{\Sigma_2(\bk,z)}{z - \omega} \; , 
\label{eq:KK}
\end{equation}
 with $\omega_0$ being a cutoff energy.
Substituting Eq.~(\ref{eq:eq14}) into Eq.~(\ref{eq:KK}) for the zero doping, 
 we obtain 
  $\Sigma_1(k,\omega) = b \omega$ with  $b = (2C/\pi)(T vk)/(\eno \omega_0)$
 and  $C= 1.25 \times 10^{-2}\lambda/\lambda_0$ for $\omega/\omega_0 \ll 1$.
 Since $ (i \omega_n \rightarrow)\; \omega$   
   in Eq.~(\ref{eq:green_function}) is replaced by  $(1+b) \omega$ with
 $b \ll 1$,
 the real part $\Sigma_1$ can be  neglected for small doping.

Finally,  we compare the present resistivity $\rho (T)$ with 
  that of the experimental one of the organic conductor
  showing almost $T$-independent $\rho (T)$ \cite{Tajima2007_EPL},  
  where  $\rho(0) > \rho (T)$ at zero doping. 
 There is a minmum $ \rho(T_{\rm min})/\rho (0) \simeq 0.3$ 
    at $T=T_{\rm min} \simeq 0.0007$.  
 When $\eno \simeq 0.0002$ and 
 $\lambda/\lambda_0 = 5$ (i.e., $\lambda/2\pi v \simeq 0.5$) are 
 assumed, 
 Eq.~(\ref{Eq_app_sigma1}) leads to  
 $\rho (0) / \rho(T_{\rm min}) \simeq 0.26$ 
   and $T_{\rm min} \simeq 0.0014$, 
 which agrees  within a factor 2. 
We note  tilting of the Dirac cone 
 in the organic conductor \cite{Kobayashi2007}, 
 which is  ignored in  the first term of Eq.~(\ref{eq:H_0}).
The tilting   gives anisotropic conductivity, \cite{Suzumura2014}, 
 but is irrelevant to   
  the almost $T$-independent $\sigma(T)$, \cite{Tajima2007_EPL}
  which is determined by  the competition 
    between  the  phonon and  the DOS of the Dirac cone. 
Further, we comment on the effect of the short-range correlation, which 
 gives rise to the insulating state and also 
 the fluctuation  close to the boundary. \cite{Kajita_JPSJ2014,Liu2016}
 Since the conductivity was measured at a pressure 
   being  much higher than that of the boundary, \cite{Tajima2007_EPL} 
  such a  correlation effect is small and the phonon scattering 
   becomes  dominant at finite temperatures.

In summary, we examined the effect of  phonon  scattering 
 on the $T$ dependence of $\sigma(T)$ 
 for both zero ($\mu_0=0$) and  finite  
($\mu_0 \not= 0$) dopings. 
  We found  $\sigma(T)$ with almost constant behavior 
   which arises from  the competition 
    between the enhancement by the Dirac cone 
     and the suppression by the acoustic  phonon. 
 Such unconventional $T$ dependence comes from the fact that  the momentum space of the phonon  scattering is strongly reduced  by  the Dirac cone.

\acknowledgments
One of the authors (Y.S.) thanks A. Kobayashi 
 for useful discussions  
  at the early stage of the present work, and  R. Kato for valuable 
 comments. 
This work was supported  by JSPS KAKENHI Grant No. JP15H02108.


\begin{thebibliography}{0}

\bibitem{Novoselov2005_Nature438}
K. S. Novoselov, A. K. Geim, S. V. Morozov, D. Jiang, M. I. Katsnelson,  
I. V. Grigorieva, S. V. Dubonos, and A. A. Firsov, 
Nature {\bf 438}, 197 (2005).

\bibitem{Berry1984}
M. V. Berry, 
Proc. R. Soc. London, Ser. A \textbf{392} (1984) 45.


\bibitem{Ando2005_JPSJ} 
T. Ando, 
 J. Phys. Soc. Jpn. \textbf{74}, 777 (2005).

\bibitem{Fradkin1986}
E. Fradkin,
 Phys.Rev. B  \textbf{33}, 3263 (1986).


\bibitem{Ziegler2007}
K. Ziegler,
  Phys.Rev. B  \textbf{75}, 233407 (2007). 

\bibitem{Castro2009_RMP}
A. H. Castro Neto, F. Guinea, N. M. R. Peres, K. S. Novoselov, and A. K. Geim, 
Rev. Mod. Phys.  \textbf{81}, 109 (2009).

\bibitem{Kajita_JPSJ2014} 
K. Kajita, Y. Nishio, N. Tajima, Y. Suzumura, and A. Kobayashi, 
 J. Phys. Soc. Jpn. \textbf{83},  072002 (2014).

\bibitem{Kato_JACS} 
R. Kato, H.B. Cui, T. Tsumuraya, T. Miyazaki, and Y. Suzumura,
J. Am. Chem. Soc. \textbf{139}, 1770 (2017). 

\bibitem{Katayama2006_JPSJ75} 
S. Katayama, A. Kobayashi, and Y. Suzumura, 
 J. Phys. Soc. Jpn. \textbf{75},  054705 (2006).

\bibitem{Kondo2005} 
R. Kondo, S. Kagoshima, and J. Harada, Rev. Sci. Instrum. \textbf{76},
 093902 (2005).

\bibitem{Kato2017_JPSJ} 
R. Kato and Y. Suzumura,
 J. Phys. Soc. Jpn. \textbf{86},  064705 (2017). 

\bibitem{Ando1998} 
N. H. Shon and T. Ando,
 J. Phys. Soc. Jpn. \textbf{67}, 2421 (1998). 

\bibitem{Katayama2006_cond} 
S. Katayama, A. Kobayashi, and Y. Suzumura, 
 J. Phys. Soc. Jpn. \textbf{75},  023708 (2006).


\bibitem{Kobayashi2008} 
A. Kobayashi, Y. Suzumura, and H. Fukuyama,
 J. Phys. Soc. Jpn. \textbf{77}, 064718 (2008). 

\bibitem{Tajima2007_EPL} 
N. Tajima, S. Sugawara, M. Tamura, R. Kato, Y. Nishio, and K. Kajita,
EPL, \textbf{80},  47002 (2007). 


\bibitem{Sarma2008} 
E. H. Hwang and S. Das Sarma, 
 Phys. Rev. B \textbf{77}, 115449 (2008). 


\bibitem{Sarma2011} 
S. Das Sarma, S. Adam, and E. H. Hwang,
 Rev. Mod. Phys. \textbf{83}, 407 (2011). 



\bibitem{Frohlich}
 H. Fr\"{o}hlich, Proc. Phys. Soc. A{\bf 223}, 296 (1954). 

\bibitem{Kobayashi2007} 
A. Kobayashi, S. Katayama, Y. Suzumura, and H. Fukuyama,
 J. Phys. Soc. Jpn. \textbf{76}, 034711 (2007). 

\bibitem{Holstein1964} 
T. Holstein, 
 Ann.Phys. \textbf{29}, 410 (1964).

\bibitem{Abrikosov} 
 A.A. Abrikosov, L.P. Gorkov, and I.E. Dzyaloshinskii,
{\it Methos of Quantum Field Theory in Statistical Physics},
(Prentice-Hall, Englewood Cliffs, NJ, 1963).
 
\bibitem{Katayama_EPJ} 
S. Katayama, A. Kobayashi, and Y. Suzumura, 
 Eur. Phys. J. B \textbf{67}, 139 (2009).

\bibitem{note} 
Diagramatically, the vertex correction appears in calculating 
 $\sigma (T)$. Here we have phenomenologially introduced  
 the factor $(1 - \cos \theta^{\pm})$ 
in the damping $\Gamma_{\pm }$ of the electron. 

\bibitem{Suzumura2014} 
Y. Suzumura, I. Proskurin, and M. Ogata,
 J. Phys. Soc. Jpn. \textbf{83}, 023701 (2014). 

\bibitem{Muller2009} 
M.  M\"{u}ller,  M. Br\"{a}uninger, B. Trauzettel,
Phys. Rev. Lett. \textbf{103},196801 (2009).

\bibitem{Liu2016} 
D. Liu, K. Ishikawa, R. Takehara, K. Miyagawa, M. Tanuma, and K. Kanoda,
Phys. Rev. Lett. \textbf{116},226401 (2016).


\end{thebibliography}
\end{document}